\documentclass[floatfix, prb, twocolumn, showpacs]{revtex4}

\usepackage{graphicx}

\usepackage{dcolumn}

\usepackage{bm}

\begin{document}

\preprint{\today}
\title{Far-infrared spectroscopy of spin excitations 
and Dzyaloshinskii-Moriya interactions in a Shastry-Sutherland
compound SrCu$_2$(BO$_3$)$_2$
}
\author{T.~R\~o\~om}
\email{roomtom@kbfi.ee}
\author{D.~H{\"u}vonen}
\author{U.~Nagel}
\affiliation{National
Institute of Chemical Physics and Biophysics,
    Akadeemia tee 23, 12618 Tallinn, Estonia.}

\author{J.~Hwang}
\author{T.~Timusk}
\affiliation{Department of Physics and Astronomy,
McMaster University,
Hamilton, Ontario L8S4M1, Canada.}

\author{H.~Kageyama}
\affiliation{Department of Chemistry, 
Graduate School of Science, 
Kyoto University, Kyoto, 606-8502, 
Japan
}

\date{\today }

\begin{abstract}
We have studied spin excitation spectra 
in the Shastry-Sutherland model compound SrCu$_2$(BO$_3$)$_2$ 
in magnetic fields  using 
far-infrared Fourier spectroscopy.
The transitions from the ground singlet state to the triplet state at 24\,cm$^{-1}$ 
and to several bound triplet states are induced
by the electric field component of the  far-infrared light.
To explain the light absorption in the spin system we invoke a dynamic 
Dzyaloshinskii-Moriya (DM) mechanism where light couples to a 
phonon mode, allowing the DM interaction.
Two optical phonons couple light to the 
singlet to triplet transition in SrCu$_2$(BO$_3$)$_2$. 
One is $a$-polarized and creates an intra-dimer dynamic DM along the $c$ axis. 
The other is $c$-polarized and creates an intra-dimer dynamic DM interaction,
it is in the $(ab)$ plane and perpendicular to the dimer axis.
Singlet levels at 21.5 and 28.6\,cm$^{-1}$  anti-cross
with the first triplet as is seen in  far-infrared spectra.
We used a cluster of two dimers with a periodic boundary condition to 
perform a model calculation with scaled intra- and inter-dimer exchange interactions.
Two static DM interactions are sufficient to 
describe the observed triplet state spectra.
The static inter-dimer DM in the c-direction $d_1=0.7$\,cm$^{-1}$
splits the triplet state sub-levels in zero field [C\'{e}pas et~al., Phys. Rev. Lett.
\textbf{87}, 167205 (2001)].
The static intra-dimer DM in the $(ab)$ plane 
(perpendicular to the dimer axis) $d_2=1.8$\,cm$^{-1}$,
allowed by the buckling of CuBO$_3$ planes, 
couples the triplet state to the 28.6\,cm$^{-1}$ singlet
as is seen from the avoided crossing.

\end{abstract}


\maketitle

\section{Introduction}
In spin systems with a ground singlet state and excited triplet state 
the energy gap between the singlet and the triplet can be tuned with 
an external magnetic field.
In SrCu$_2$(BO$_3$)$_2$ it was discovered that in magnetic fields above 22T,
where the spin gap is expected to close, several magnetization plateaus appear.\cite{Kageyama1999}
At magnetization plateaus the triplets form a pattern
which breaks the translational symmetry of the crystal structure.\cite{Kodama2002}
The heavy mass of the triplet excitations arising from an almost flat dispersion 
of energy on momentum\cite{Kageyama2000}
favors the build-up of magnetic superstructures.
Below the critical field SrCu$_2$(BO$_3$)$_2$ has a ground state
described first by Shastry and Sutherland.\cite{Shastry1981}

SrCu$_2$(BO$_3$)$_2$ consists of planes of CuBO$_3$  and Sr atoms between 
the planes.
Cu$^{2+}$ spins (S=1/2) form Cu-Cu dimers arranged into orthogonal dimer network.
SrCu$_2$(BO$_3$)$_2$  is an experimental realization of a Shastry-Sutherland
model.\cite{Shastry1981}
In the model there is an anti-ferromagnetic intra-dimer exchange coupling $j_1$ and 
inter-dimer coupling $j_2$ between spins 
on the nearest-neighbor dimers (Fig.\,\ref{Model2dimersStaticW}).
In the limit of $\alpha\equiv j_2/j_1=0$ the problem reduces to that of  isolated
dimers where the ground state is the product of singlet states and 
the first excited  triplet state is at energy $\Delta_T=j_1$ above the ground 
state, where $\Delta_T$ is the energy per dimer.
Shastry and Sutherland  showed that for $0<\alpha\le 0.5$ singlets on all dimers 
is  an exact  ground state too. 
The exactness of the ground state and the heavy mass of triplet excitations is 
the consequence of frustration originating from the special geometry of the dimer lattice
in the Shastry-Sutherland model
where the  bonds on neighboring dimers are orthogonal.
Later on it has been shown that singlets on all dimers 
is the exact  ground state for a larger 
range of $\alpha$ up to the quantum critical  point $\alpha_c \approx  0.7$.
At the quantum critical point the spin gap vanishes 
and a long-range anti-ferromagnetic order is established.
Different theoretical approaches have been used to calculate 
$\alpha_c$ (see Ref.\cite{Miyahara2003} for review).
It is possible that between the exact singlet ground state and 
the anti-ferromagnetic state in certain range of $\alpha$ 
other gapped spin states 
exist.\cite{Knetter2000,Koga2000,Zheng2001a,Chung2001,Lauchli2002,Munehisa2004}

The singlet-triplet gap in SrCu$_2$(BO$_3$)$_2$, $\Delta_T=24$\,cm$^{-1}$,
has been measured directly by several experimental techniques: 
inelastic neutron scattering\cite{Kageyama2000,Cepas2001}, 
electron spin resonance\cite{Nojiri1999,Nojiri2003} (ESR), 
Raman scattering\cite{Lemmens2000}, 
and far-infrared (FIR) spectroscopy.\cite{Room2000}
Additional information  besides $\Delta_T$ is needed 
to determine the exchange parameters of SrCu$_2$(BO$_3$)$_2$.
The dispersion of the triplet excitation is not informative because of 
its flatness\cite{Kageyama2000}, but positions of other excited states or 
the temperature dependence of thermodynamic parameters can be used
for determining the exchange parameters.
Miyahara and Ueda\cite{Miyahara2003}  found $j_1=59$\,cm$^{-1}$ and 
$\alpha=0.635$.  
They added an interlayer coupling $j_3=0.09j_1$ to the model
to obtain a better fit of the magnetization $T$-\,dependence 
above the critical temperature $k_BT>\Delta_T$. 
Based on the analysis of excitation spectra\cite{Knetter2004}
$j_1=50$\,cm$^{-1}$ and $\alpha=0.603$ were proposed.
Such scattering of parameters could be 
either due to the incomplete model or due to the approximations made
in theoretical calculations.
SrCu$_2$(BO$_3$)$_2$ is near to the quantum critical point $\alpha_c$ 
where the energy levels of the spin system are sensitive
to the choice of $j_1$ and  $\alpha$.
 A singlet level 
in the spin gap at 21\,cm$^{-1}$ found in the ESR spectra\cite{Nojiri2003} 
may help to find proper parameters for the model.

Interactions other  than  inter- and intra-dimer exchange coupling
can spoil the exactness of the ground state.
This is  important in  high magnetic fields where 
the triplet state becomes degenerate with the ground singlet state.
At this critical field even a weak interaction between the singlet and the 
triplet state mixes the two states completely. 
The singlet and triplet state anti-crossing effects were seen in the 
high field ESR experiments.\cite{Nojiri2003}
A possible anti-symmetric interaction which couples the singlet and the 
triplet states is the DM interaction.
An intra-dimer DM is allowed by symmetry but its strength is not known 
below room temperature.
Above room temperature $d_2=2.5$\,cm$^{-1}$ has been
estimated from the ESR linewidth.\cite{Zorko2003b}
The inter-dimer DM interaction, $d_1=1.5$\,cm$^{-1}$,   
perpendicular to the dimer planes\cite{Cepas2001} 
partially lifts  the degeneracy  of the triplet state
but does not couple the  triplet state to the singlet state.
The effect of DM interactions on the magnetic dipole active ESR transitions 
in SrCu$_2$(BO$_3$)$_2$ was investigated theoretically in Ref.\cite{Miyashita2003}.

Lattice distortions, static or dynamic,  are important in SrCu$_2$(BO$_3$)$_2$ 
since they lower the crystal symmetry and allow magnetic interactions
which are otherwise forbidden in a more symmetric environment.
SrCu$_2$(BO$_3$)$_2$ has a structural phase transition at 395\,K\cite{Sparta2001}
that induces a buckling of CuBO$_3$ planes in the low $T$ phase.
As the phase transition point is approached from below the Raman-active
62\,cm$^{-1}$ optical phonon mode softens.\cite{Choi2003}
Acoustic phonon modes have spin-phonon coupling at magnetization 
plateaus.\cite{Wolf2001}
It has been proposed that a spin superstructure at 1/8 plateau
observed by nuclear magnetic resonance  at 35\,mK 
is stabilized by a lattice distortion.\cite{Kodama2002} 
Instantaneous breaking of lattice symmetry by an optical phonon
allows electric dipole active singlet-triplet transitions\cite{Cepas2004PRB}
that explains FIR polarized absorption spectra 
in SrCu$_2$(BO$_3$)$_2$.\cite{Room2000} 

Our aim is to find out which additional interactions are required
to the Shastry-Sutherland model that add triplet corrections to the ground state.
For that we do FIR  absorption measurements 
with polarized light in magnetic field
and compare the absorption line frequencies and intensities 
with values calculated with a two dimer model including the dynamic
DM effect.
The important information is in the polarization and magnetic field dependence 
of the FIR absorption lines and in the avoided  crossing effects.

We studied single crystals of  SrCu$_2$(BO$_3$)$_2$, Ref.\,\cite{Kageyama1999a}.
	The first sample consisted of two pieces 0.65\,mm thick in a-direction
with the total area of 12\,mm$^2$ in the $(ac)$ plane.
	The second sample was 0.6\,mm thick in the c-direction and had an 
area of 11.5\,mm$^2$ in the $(ab)$ plane.
The experimental details are described in Ref. \cite{Room2000,Room2004}.

\section{Results and discussion}

\subsection{FIR spectra and electric dipole transitions }\label{ElectricDipole}

\squeezetable
\begin{table}[tb]
\caption{\label{AllLines}
Singlet and triplet excitations observed in the FIR spectra 
at 4.4\,K in the order of
increasing zero field energies $\hbar\omega_0$ (in cm$^{-1}$ units).
When a line is visible in two $\mathbf{E}_1$ polarizations, 
both are indicated. 
The corresponding $\mathbf{H}_1$ polarizations are also indicated.
$S$ and $T$ label the singlet and triplet states;
$+$  ($-$) denotes levels which energy increases (decreases) with $\mathbf{B}_0$ 
and $0$ indicates levels where the energy stays constant; 
$g_{a}$ and $g_{c}$ are the $g$-factors
with $\mathbf{B}_0\,\parallel\,\mathbf{a}$  and $\mathbf{B}_0\,\parallel\,\mathbf{c}$, respectively.
The labeling of $T_0$ levels is shown in Fig.\,\ref{EparAEigenProb}. 
The zero field intensities  $A_0$ (in \,cm$^{-2}$ units)  of  $T_0$  ($\star$) are described 
in the text and in figures
\ref{BacEcLevelsProbab}, \ref{EparAEigenProb}, and \ref{EaBbyEigenProb}.
High energy excitations are labelled by their energies.
}
\begin{tabular}{l|d|d|lldddd}
& & & & & & &  \\ 
Label& \mathbf{E}_1 & \mathbf{H}_1 & $A_0$ & $\hbar\omega_{\,0}$ &   & g_{a} &  g_{c}    \\
\hline & & & & & &  \\ 
$ S_1$ & a    & c & $ $ & 21.50 & \pm 0.03 &   &   \\
$T_{0m}(\pm)$  & a, c & c, a &   $\star$   & 22.72 & \pm 0.05 & 1.988 & 2.219 \\ 
$T_{0p,m}(0)$ & a, c & c, a &   $\star$   & 24.11 & \pm 0.05 &   &   \\
$T_{0p}(\pm)$ & a, c & c, a &   $\star$   & 25.51 & \pm 0.05 & 1.988 & 2.219 \\ 
$S_2$ & a, c  & c, a & $ $ & 28.57 & \pm 0.03 &   &   \\
$T_1(\pm)$      & c    & a & 0.3 $\pm$ 0.2  & 37.49 & \pm 0.03 & 1.996 & 2.264 \\ 
$T_1(\pm)$      & a    & b, c& 0.9 $\pm$ 0.2 & 37.51 & \pm 0.04 & 2.001 & 2.23 \\ 
$T_1(0)  $ & a 	& c & 0.9 $\pm$ 0.2 & 37.69 & \pm 0.09 &   &   \\ 
$T_{38.7}(\pm)$ & a 	& c & $ $    & 38.74 & \pm 0.03 & 2.026 &   \\ 
$T_{38.7}(0)$ & c 	& a & $ $    & 38.70 & \pm 0.15 &  &   \\
$T_{39.1}(\pm)$ & c 	& a & $ $    & 39.08 & \pm 0.15 & 2.067 & 2.29 \\ 
$S_{39.7}$ & a 	& c & 0.19 $\pm$ 0.05 & 39.71 & \pm 0.04 &   &   \\ 
$T_{40.5}(\pm)$ & a 	& c & $ $     & 40.45 & \pm 0.03 & 1.97 &   \\ 
$T_{40.7} (\pm) $ & c 	& a & 0.2 $\pm$ 0.1 & 40.67 & \pm 0.03 &   & 2.243 \\ 
$T_{40.7}(0)$ & a, c & c, a & 0.2 $\pm$ 0.1 & 40.70 & \pm 0.16 &   &   \\ 
$T_{41.1}( ) $ & c 	& a & 0.4 $\pm$ 0.1 & 41.11 & \pm 0.13 & 2.10 &   \\ 
$T_{42.7}(+) $ & a 	& c & 0.2 $\pm$ 0.1 & 42.7  & \pm 0.2  &   & 2.25 \\ 
$S_{43}$   & a & b, c & 2.6 $\pm$ 0.3 & 43.00 & \pm 0.16 &   &   \\ 
$T_{43.5}(\pm) $ & c 	& a & 0.2 $\pm$ 0.1 & 43.54 & \pm 0.03 &   & 2.31 \\ 
$S_{44.7}$ & c 	& a & $ $    & 44.7  & \pm 0.4  &   &   \\ 
$S_{47.0}$ & c 	& a & $ $    & 47.04 & \pm 0.04 &   &   \\ 
$T_{48.2}(\pm) $ & c 	& a & 0.04 $\pm$ 0.02 & 48.21 & \pm 0.09 &   & 2.27 \\ 
$S_{52.3}$ & a 	& b, c & 86 $\pm$ 14    & 52.24 & \pm 0.08 &   &   \\ 
$S_{53.5}$ & a 	& b, c & 24 $\pm$ 3     & 53.44 & \pm 0.07 &   &    
\end{tabular}
\end{table}

\begin{figure}[tb]
    \includegraphics[width=8.6cm]{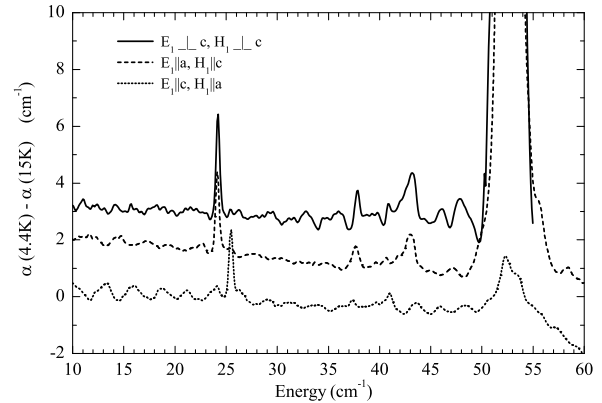}

\caption{Differential absorption
in $\mathbf{E}_1\!\perp  \!\mathbf{c}$ (two upper curves)
and $\mathbf{E}_1\!\parallel  \!\mathbf{c}$ (lower curve) polarization.
Spectra have been off-set in vertical direction.
}
\label{EaEc4K4spec}
\end{figure}

As the result of the polarization sensitive measurement of FIR spectra we have 
identified that the main resonances in the spectra are electric dipole transitions,
rather than being magnetic dipole transitions.
In Fig.~\ref{EaEc4K4spec} differential absorption spectra at 4.4~K relative to 15~K, 
are displayed. The strong absorption lines at 52.3 and 53.5\,cm$^{-1}$ were identified\cite{Room2000} 
as electric dipole transitions, that are active in 
$\mathbf{E}_1\,\parallel\,\mathbf{a}$ polarization.
We see  the same for 
the 43.0\,cm$^{-1}$  singlet and $T_0$ and $T_1$ triplets (see Table\,\ref{AllLines}) at
24.2 and 37.5 cm$^{-1}$, respectively, which are present in the spectra
measured with $\mathbf{E}_1\,\parallel\,\mathbf{a}$ regardless of $\mathbf{H}_1$
being perpendicular to the $c$ axis or parallel to it. 
The lines are missing in $\mathbf{E}_1\,\parallel\,\mathbf{c}$
polarization\cite{misalignement}, instead a new line appears at 25.5\,cm$^{-1}$, which is identified 
as another component of the triplet $T_0$.

\begin{figure}[tb]
    \includegraphics[width=8.6cm]{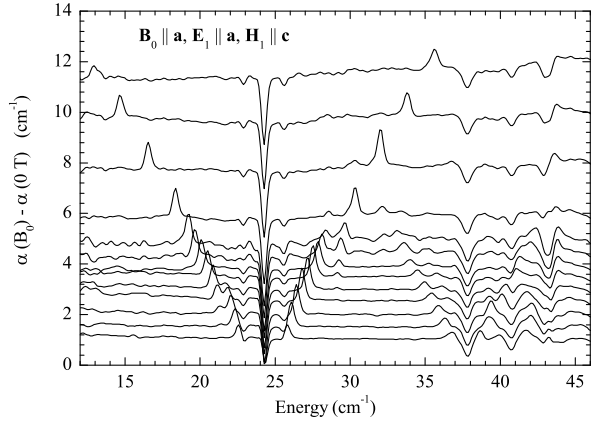}
\caption{
Differential absorption spectra in magnetic field $\mathbf{B}_0 \! \parallel \! \mathbf{a}$
at 4.4\,K.
Vertical offset equals to the magnetic field value in Tesla.}
\label{BaEaSpec}
\end{figure}

The triplets are split by the magnetic field $\mathbf{B}_0$. 
Differential absorption spectra in 
$\mathbf{E}_1\!\parallel\!\mathbf{a}$
polarization for one magnetic field direction,
$\mathbf{B}_0\!\parallel\!\mathbf{a}$, measured relative to the zero field,
are displayed in Fig.\, \ref{BaEaSpec}. We see an anti-crossing of the 
$T_{0m}(-)$ level with the singlet $S_1$ at 21.5~cm$^{-1}$ 
and an anti-crossing of the
$T_{0p}(+)$ level with the singlet $S_2$ at 28.6~cm$^{-1}$.	
All the peaks in the measured spectra 
in different light polarizations and $\mathbf{B}_0$ directions
were fitted with Lorentzians. 
The results are summarized in Table\,\ref{AllLines}
and displayed in Figures \ref{BacEcLevelsProbab}, \ref{EparAEigenProb}, and \ref{EaBbyEigenProb}. 
The states above 38\,cm$^{-1}$ are labelled by their zero field frequencies.
The magnetic field independent energy levels are labelled as singlets
with the exception of those in the middle of the triplet levels
$T(\pm)$.

\subsection{Dynamic Dzyaloshinskii-Moriya mechanism and optical transitions: two dimers}

The hamiltonian for a spin pair with exchange coupling $j$ and DM interaction
$\mathbf{d}$ on the bond connecting spins $k$ and $l$ reads:
\begin{eqnarray}
H_{stat}^{kl}& =& (j-\frac{| \mathbf{d} |^2}{4j}) \mathbf{S}_k \cdot  \mathbf{S}_l +
\frac{1}{2j}\mathbf{S}_k \cdot \mathbf{d}\mathbf{d} \cdot \mathbf{S}_l \nonumber \\
& + &\mathbf{d} \cdot  [\mathbf{S}_k \times   \mathbf{S}_l]
+ g\mu _B \mathbf{B}_0 \cdot (\mathbf{S}_k +\mathbf{S}_l).
\label{pair_static}
\end{eqnarray}
Here we included   Shekhtman corrections\cite{Shekhtman1992,Shekhtman1993}
which are quadratic in $\mathbf{d}$ (see also\cite{Room2004}).
The last term is the Zeeman energy of spins in the magnetic field $\mathbf{B}_0$
where $g$ is the electron spin g-factor and $\mu_B$ is the Bohr magneton. 

The formalism to introduce the spin-phonon coupling is similar to one 
used in Ref.\cite{Cepas2004PRB,Room2004}.
We are interested in  singlet to triplet transitions.
Therefore the relevant term is the anti-symmetric  DM interaction 
$ \mathbf{d}(Q) \cdot  [\mathbf{S}_k \times   \mathbf{S}_l] $
which couples the singlet to the triplet state.
We expand the DM vector $ \mathbf{d} (Q)$ into a power series 
of the lattice normal coordinate $Q$
\begin{equation}
\mathbf{d}(Q) =  \mathbf{d}(0) + 
\frac{\partial  \mathbf{d}}{\partial Q}\left| \right._{Q=0}Q + \ldots., 
\label{series}
\end{equation}
where $\mathbf{d}(0)\equiv \mathbf{d} $ is the static DM interaction 
in (\ref{pair_static}).
We keep  terms linear in $Q$.
The full hamiltonian for a spin pair including the phonons is
\begin{equation}
H^{kl}  = H_{stat}^{kl}  + \hbar \, \omega_p \, a^\dagger a + 
  q(a^\dagger + a)\,\mathbf{d}_Q  \cdot  [\mathbf{S}_k \times   \mathbf{S}_l],
\label{dynamic0order}
\end{equation} 
where $ \mathbf{d}_Q \equiv  \frac{\partial  \mathbf{d}}{\partial Q}\left| \right._{Q=0} $.
The lattice normal coordinate $ Q$ is presented in terms of phonon
creation and annihilation operators $a^\dagger$ and $a$, $ Q = q(a^\dagger + a) $,
where $q$ is the transformation coefficient and $\omega_p$ is the phonon frequency.
The spin-phonon coupling term in (\ref{dynamic0order}) is linear in $a^\dagger$ and $a$.
Therefore the phonon states with the occupation numbers $n$ and $n^\prime$ 
are coupled where $n^\prime = n\pm 1$.
We will consider only two phonon states $| 0 \rangle$ and $| 1 \rangle$, 
which is justified when $k_BT\ll \hbar\omega_p$.

The normal coordinate $Q$ 
in the dynamic DM singlet to triplet optical transition mechanism 
belongs to an optical phonon.
Electric dipole coupling between a phonon and light 
in the long wavelength limit is 
\begin{equation}
V  =  eQE_1
 =   e q(a^\dagger + a)E_1,
\label{eldipoleham}
\end{equation}
where $e$ is the effective charge associated with the lattice
normal coordinate $Q$. 
Here we assumed $\mathbf{E}_1 \parallel \mathbf{Q}$
and dropped the time dependence of $V$.
Once the eigenstates of (\ref{dynamic0order}) are known
the optical transition probability between the ground state  
$ |\phi \rangle $ and the excited state $ |\phi^\prime \rangle $
is calculated as $ I=|\langle \phi^\prime |V| \phi \rangle|^2 $.

\begin{figure}[bt]
\includegraphics[width=8.6cm]{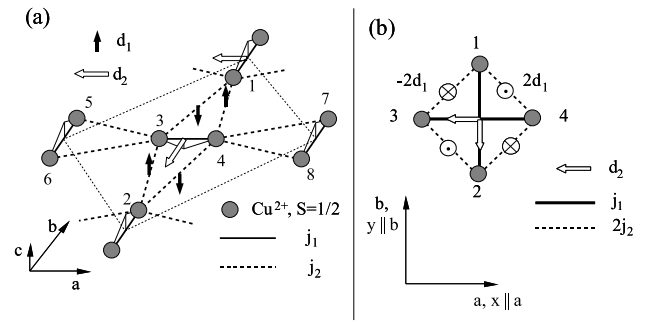}
\caption{Cluster with two dimers $(1,2)$ and $(3,4)$. (a)  Dimer $(3,4)$
and four nearest-neighbor dimers. The thin dashed line shows the two dimer cluster
boundary. Thin solid lines show the distortion of Cu-Cu superexchange bonds due to the 
buckling of Cu-O-B planes. Thick solid and dashed lines are the inter- and intra-dimer
superexchange constants $j_1$ and  $j_2$; 
inter-dimer DM vectors ($\mathbf{d}_1$, solid arrow)
are in the $c$ direction and intra-dimer DM vectors ($\mathbf{d}_2$, empty arrow)
in the $(ab)$ plane along $a$ and $b$ axis.
(b)  The two dimer model after the periodic boundary condition has been applied;
inter-dimer interactions have doubled.
}
\label{Model2dimersStaticW}
\end{figure}

To calculate optical transitions in SrCu$_2$(BO$_3$)$_2$ we use a two dimer model
depicted in Fig.\,\ref{Model2dimersStaticW}.
In this model intra-dimer and inter-dimer superexchange interactions
$j_1$ and $j_2$ are considered.
The inter-dimer static DM vector $\mathbf{d}_1$ is along the $c$ axis 
and alternates from bond to bond.
The intra-dimer static DM vector $\mathbf{d}_2$ exists due to the buckling 
of Cu-O-B planes.\cite{Sparta2001}
The direction of DM vectors is defined by the right hand rule
where the path is along the Cu-O-Cu bond (for $\mathbf{d}_1$ Cu-O-B-O-Cu)
in the direction of increasing spin index $k$.
In the vector product $ \mathbf{S}_{\,k} \times \mathbf{S}_{\,l}  $ 
the spin with a smaller index is on the left, $k<l$.
When a periodic boundary condition is applied to the two dimer cluster,
bounded by a box drawn with a thin dashed line in Fig.\,\ref{Model2dimersStaticW}(a),
an effective spin model is obtained where the inter-dimer interactions are doubled,
Fig.\,\ref{Model2dimersStaticW}(b).
The doubling is necessary to conserve the number of next-nearest-neighbor bonds,
which is four.

\begin{figure}[tb]
\includegraphics[width=8.6cm]{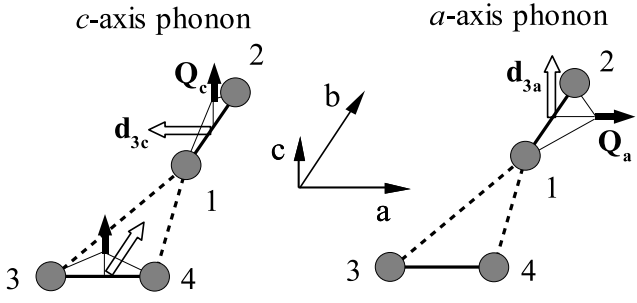}
\caption{Intra-dimer dynamic DM interactions. 
A lattice  distortion with the normal coordinate $\mathbf{Q}$ (solid arrow) creates 
an intra-dimer DM interaction  $\mathbf{d}_3$ (empty arrow).
The $c$-axis phonon creates a dynamic DM interaction on both dimers
while the $a$-axis phonon affects the dimer $(1,2)$ only.
}
\label{Model2dimersDynamic}
\end{figure}

The hamiltonian for the two dimer cluster  is the sum of pairwise interactions  
(\ref{dynamic0order}) where the sum runs over all the bonds in the cluster.
We will use a  basis $ | ABn \rangle $
where $A$ runs over the singlet $S$  and three triplet components 
$T_-, T_0$,  and $ T_+$ 
on the $j_1$ bond of the dimer $(1,2)$
and $B$ over the singlet and triplet  states of the dimer $(3,4)$.
$n$ is the number of phonons, 0 or 1.
The basis has 32 components.
Below we  consider $a$- and $c$-axis phonons, shown in Fig.\,\ref{Model2dimersDynamic}, 
named by the direction of their electric dipole moment.

\subsubsection{Energy levels}
The effect of the dynamic DM interaction on the position of  
energy levels is small because  we  take $\hbar \omega_p = 100 $\,cm$^{-1}$
that is substantially larger than the singlet-triplet gap.
We use this value since there are no optical phonons with substantial spectral weight 
below 100\,cm$^{-1}$ as our transmission measurements show.
The energy spectrum can be analyzed separately from the dynamic DM effect
because of the high phonon energy. 
The calculated energy levels are the same in Fig.\,\ref{BacEcLevelsProbab} 
and \ref{EparAEigenProb}.
In these figures only the zero phonon levels of the triplet $T_0$  
and $S_2$ are shown.
The levels with one excited phonon are off-set by $\hbar \omega_p$
to higher energies and are not shown.

In a two dimer system two singlets, two triplets, and a quintet are present.
The  ground state is a product of  singlets 
$ |SS \rangle $. 
The first triplet is a linear combination of 
$ | ST \rangle $  and $ | TS \rangle $.
In the two dimer model the singlet-triplet splitting
is not renormalized by the inter-dimer coupling $j_2$
and the energy of the triplet excitation is $E_{T0}=j_1$.
The second singlet, a bound state of two triplets, is at $E_{S1}=2j_1-2(2j_2)$.
To stress the fact that in the two dimer model with a periodic boundary condition
the inter-dimer bonds are effectively doubled,
we write  $2j_2$ explicitly.
There are two other bound states of two triplets,
a triplet at $E_{T1}=2j_1-(2j_2)$ and a quintet at 
$E_{Q}=2j_1+(2j_2)$.
These energies and the ground state wavefunction 
are slightly changed by the static DM interactions $d_1$ and $d_2$.
The spin  states   
 $ | ST_i \rangle $  and $ | T_iS \rangle $
are strongly mixed by the inter-dimer $d_1$ since they are degenerate
in any field.

The states are labelled in Fig.\,\ref{EparAEigenProb}.
The following parameters were used to fit the energy spectra plotted in 
Fig.\,\ref{BacEcLevelsProbab} and \ref{EparAEigenProb}.
The energy of one-triplet sublevels $T_{0m}(0)$ and $T_{0p}(0)$
gives us $j_1=24.0$\,cm$^{-1}$.
To get the   singlet $S_2$ at 28.6\,cm$^{-1}$
we use $2j_2=9.8$\,cm$^{-1}$.
Triplet levels are split in zero field
by $2d_1=1.4$\,cm$^{-1}$.
The intra-dimer $d_2=1.8$\,cm$^{-1}$ induces an avoided crossing of 
$T_{0p}(+)$ and $S_2$.
In a simplified picture the one-triplet excitation is 
the $ | ST \rangle $ (or $ | TS \rangle $ ) state
and the excited  singlet is $ | TT \rangle $.	
$d_2$ ``flips'' the singlet to the triplet state on 
one of the dimers 
and thus couples $T_{0p}(+)$ to $S_2$.

\subsubsection{$c$-axis phonon}
The optical $c$-axis phonon bends the Cu-O-Cu bond in the c-direction.
We assume that the bending action of the phonon 
is the same on both dimers, Fig.\,\ref{Model2dimersDynamic}.
As a result the dynamic DM interaction on the dimer (1,2) is 
$q_c\mathbf{d}_{Q_c} \equiv\mathbf{d}_{3c}=(-d_{3c},0,0)$ 
and on the dimer (3,4) $\mathbf{d}_{3c}=(0,d_{3c},0)$; 
the orientation of the Cartesian coordinates is the same 
as in Fig.\,\ref{Model2dimersStaticW}\,(b).
The calculated and the measured transition probabilities as a function
of magnetic field are plotted in Fig.\,\ref{BacEcLevelsProbab}\,(b, d) 
for two field orientations.
In zero field a line at 25.5\,cm$^{-1}$ is present.
The area of this line is the only scaling parameter
between the theory and the experiment.
Note that the transition to the triplet level, 
which anti-crosses with $S_2$,
is optically active when $\mathbf{B}_0\parallel \mathbf{c}$.
When $\mathbf{B}_0\parallel \mathbf{a}$ there is no crossing 
for the optically active triplet level.

The overall agreement between the theory and the experiment is good.
There is a disagreement between the intensities of the middle
and lower triplet components in the theory and in the experiment, 
Fig.\,\ref{BacEcLevelsProbab}(d).
In the theory the intensity of the middle component is approximately
three times as strong as the lower component 
while in the experiment they are equal.
We tried several changes in our model to make the intensities of the two 
triplet components more equal and none of them helped.
These unfruitful changes were the shift of the phonon frequency,
a small out-of-plane component of $\mathbf{B}_0$
and an in-plane component of the inter-dimer DM vector $\mathbf{d}_1$.

\subsubsection{$a$-axis phonon}

The optical $a$-axis phonon bends the Cu-O-Cu bond in the a-direction
and creates a dynamic DM interaction in the c-direction, 
Fig.\,\ref{Model2dimersDynamic}.
If we choose  $\mathbf{E}_1 \parallel \mathbf{a}$ 
the dynamic DM interaction is created on dimer (1,2),
$q_a\mathbf{d}_{Q_a} \equiv\mathbf{d}_{3a}=(0,0,d_{3a})$.
In general, for an arbitrary orientation of $\mathbf{E}_1 $ in the $(ab)$ plane,
both dimers will acquire a certain $\mathbf{d}_{3a} $.
For the time being we assume  $\mathbf{E}_1 \parallel \mathbf{a}$.

In zero magnetic field the transition to the central triplet component
is observed, Fig.\,\ref{EaEc4K4spec}.
As the $\mathbf{B}_0 \parallel \mathbf{c}$ field is turned on,
Fig.\,\ref{EparAEigenProb}\,b, 
the central line, being a sum of two overlapping transitions, 
conserves its intensity.
The experimentally observed drop in intensity 
with increasing field is a $T$ effect.
At 1.8\,K  (18\,T field) the intensity is recovered.
Besides the strong central line there are in zero field
two  sidepeaks ten times weaker at 22.7 and 25.5\,cm$^{-1}$ 
corresponding to transitions to the twice degenerate states 
$T_{0m}(\pm)$ and $T_{0p}(\pm)$.
The dynamic DM interactions due to the $a$ axis
and $c$-axis phonons in this $\mathbf{B}_0$ orientation 
give zero intensity for the sidepeaks.
The detailed analysis of the mechanism causing
these weak transitions is difficult because in other
polarizations and field orientations 
stronger mechanisms are prevailing.
The sidepeaks split in the magnetic field and
an avoided crossing with $S_1$ and $S_2$
is seen in the experiment.

\begin{figure}[tb]
    \includegraphics[width=8.6cm]{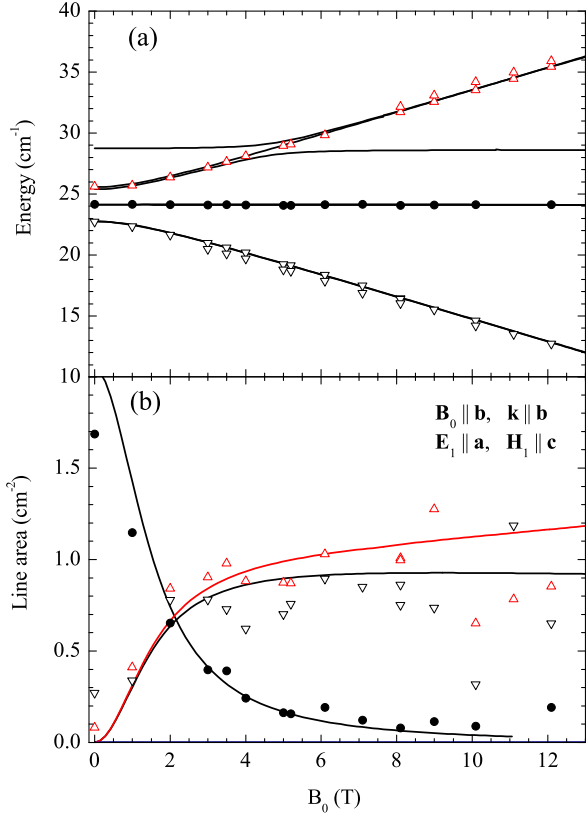}
\caption{(color online).
Line positions (a)  and line areas (b) in $\mathbf{E}_1 \! \parallel \! \mathbf{a}$  
and 
$\mathbf{B}_0 \! \parallel \! \mathbf{b}$ configuration at 4.4K.
The lines are results of the calculation based on the two dimer model and dynamic 
DM interaction. 
The additional splitting of triplet components (triangles) is caused by the magnetic 
field $\mathbf{B}_0$
being misaligned by $9^\circ$ out of the $(ab)$ plane.\cite{misalignement}
In panel (b) the line area (triangles up or triangles down)
is a sum of line areas of split components. 
}
\label{EaBbyEigenProb}
\end{figure}

When the magnetic field is in the $(ab)$ plane
two  cases must be considered,
$\mathbf{B}_0 \parallel \mathbf{E}_1  $ and 
$\mathbf{B}_0 \perp \mathbf{E}_1  $.
In Fig.\,\ref{EparAEigenProb}\,(c, d) 
the $\mathbf{B}_0 \parallel \mathbf{E}_1 $ case is shown.
Here are optically  active  the triplet levels which anti-cross
with the singlet states.
In $\mathbf{B}_0 \perp \mathbf{E}_1  $ field orientation,
Fig.\,\ref{EaBbyEigenProb},
the optically active triplet levels do not anti-cross 
with the singlet states.
The mutual orientation of $\mathbf{B}_0$ and $\mathbf{E}_1  $ 
is important because  $\mathbf{E}_1 \parallel \mathbf{a}$
creates $ \mathbf{d}_{3a} $ on the dimer (1,2) and
not on (3,4).
Which set of the two-fold degenerate triplet levels 
is optically active depends on
the relative orientation of $\mathbf{B}_0$ and  $ \mathbf{d}_2 $ 
on the dimer where $d_{3a}\neq 0$.
In Fig.\,\ref{EparAEigenProb} $\mathbf{B}_0 \parallel \mathbf{d}_2  $ 
and in Fig.\,\ref{EaBbyEigenProb} $\mathbf{B}_0 \perp \mathbf{d}_2  $.
An additional splitting of  $T_{0m}(\pm)$ and $T_{0p}(\pm)$
by 0.6\,cm$^{-1}$ seen in Fig.\,\ref{EaBbyEigenProb} 
is because $\mathbf{B}_0$ is out of $(ab)$ plane by $9^\circ$.

\subsection{Static and dynamic DM in SrCu$_2$(BO$_3$)$_2$}
\label{DMdiscussion}

We have shown  that the first triplet state energy spectra
are well described with two static DM interactions, 
$ \mathbf{d}_{1} $  and $ \mathbf{d}_{2} $.
The information about $ \mathbf{d}_{1} $  and $ \mathbf{d}_{2} $
is contained in the position of energy levels and in the 
FIR absorption line intensities.
The inter-dimer $ \mathbf{d}_{1} $ 
determines the magnetic field dependence of intensities
and the triplet state level energy splitting.
The intra-dimer $ \mathbf{d}_{2} $ 
determines the extent of the avoided crossing 
with $S_2$ 
and the magnetic field dependence of intensities near 
the avoided crossing points.
Over the magnetic field range of our experiment
the intensities of the singlet-triplet absorption lines
do not depend on the dynamic part of the DM interaction,
because the  phonon energies are large compared 
to the triplet state energy.
 
Other inter- and intra-dimer DM interaction components
besides $ \mathbf{d}_{1} $  and $ \mathbf{d}_{2} $
have been considered  to describe experimental 
data.\cite{Jorge2003,Zorko2003b}
These are  the in-plane component of 
the inter-dimer DM $ \mathbf{d}_{xy} $
and  the symmetry-forbidden intra-dimer DM  $ \mathbf{d}_{z} $
in the c-direction.
We included $ \mathbf{d}_{xy} $  and $ \mathbf{d}_{z} $
in the two dimer model and found that calculations
with non-zero  $ \mathbf{d}_{xy} $  and $ \mathbf{d}_{z} $
give results  contradicting with the experiment.
Our  argument, which is independent of 
whether a particular infrared transition is allowed or forbidden,
relies on the observed and calculated crossing - anti-crossing
effects between the triplet and the singlet states.

If $\mathbf{B}_0 \parallel \mathbf{c} $ and $ \mathbf{d}_{xy} \neq 0 $   
then $T_{0m}(+)$ 
would have  an avoided crossing with $S_2$
contradicting the experiment, where $T_{0p}(+)$
anti-crosses with the singlet [Fig.\,\ref{EparAEigenProb}(a)].
Also $ \mathbf{d}_{z} $ does not give any anti-crossing 
between  $S_2$ and 
$T_{0m}(+)$ or $T_{0p}(+)$.
In high field nonzero $ \mathbf{d}_{2} $ creates an avoided crossing
between the ground state $S_0$ and the  triplet branch $T_{0m}(-)$
as  observed in the experiment\cite{Nojiri2003}
while nonzero $ \mathbf{d}_{xy} $  or  $ \mathbf{d}_{z} $
do not create an avoided crossing between  $S_0$ 
and $T_{0m}(-)$ or $T_{0p}(-)$.
However, the two dimer model does not predict 
the experimentally observed\cite{Nojiri2003} avoided crossing
between $S_0$  and $T_{0p}(-)$.

In  $ \mathbf{B}_0 \parallel \mathbf{a} $ field orientation
both $ \mathbf{d}_{xy} $  and $ \mathbf{d}_{z} $ add, 
in addition to $ \mathbf{d}_{2} $,
to the avoided crossing of one of the triplet 
components with $S_2$.
The experimental data can be fitted with a single value
$ d_{2}=1.8 $\,cm$^{-1}$ in both
field orientations, $ \mathbf{B}_0 \parallel \mathbf{a} $ 
and $ \mathbf{B}_0 \parallel \mathbf{c} $.
If  $ \mathbf{d}_{xy} $  and $ \mathbf{d}_{z} $ 
were comparable in magnitude to $ \mathbf{d}_{2} $,
then the extent of avoided crossing 
would be different in 
$ \mathbf{B}_0 \parallel \mathbf{a} $ and 
$ \mathbf{B}_0 \parallel \mathbf{c} $ 
field orientations.

Our conclusion is that the dominant DM interactions  are 
$ d_1 = 0.7 $\,cm$^{-1}$ 
and $ d_2 = 1.8 $\,cm$^{-1}$. 
In  the magnetization plateau state   
the lattice parameters of SrCu$_2$(BO$_3$)$_2$ may change
due to spin-phonon coupling.\cite{Wolf2001}
Our calculation of energy levels did not account for static lattice distortions
and therefore 
we cannot make any conclusions about 
$ \mathbf{d}_{xy} $  and $ \mathbf{d}_{z} $ and the strength 
of $ \mathbf{d}_1$ and $ \mathbf{d}_2$ in high magnetic fields.

The intensity of the FIR singlet-triplet transitions depend on 
the strength of the dynamic DM and  
on the  frequency  and the oscillator strength of the phonon.  
Since the particular phonons
involved in the dynamic DM effect in SrCu$_2$(BO$_3$)$_2$ are not known
we can give only the relative strength of dynamic DM interactions.
The $a$- and $c$-polarized singlet-triplet transitions have similar 
oscillator strengths.
These are 2.0\,cm$^{-2}$ ($ \mathbf{E}_1 \parallel \mathbf{a} $ ) 
and 1.7\,cm$^{-2}$ ($ \mathbf{E}_1 \parallel \mathbf{c} $ ) 
if we compare the two lower spectra in Fig.\,\ref{EaEc4K4spec}  
which have been measured on the same sample 
by changing the direction of the light polarization.
The ratio  of the dynamic DM interactions for the two mechanisms is
$d_{3\,a} / d_{3\,c} = \sqrt{ 2 \times 2.0/1.7} = 1.5$
if we assume that $a$- and $c$-axis phonons
have equal frequencies and oscillator strengths.
The factor 2 accounts for the  $a$-axis phonon
creating a dynamic DM only on the dimer with its axis perpendicular to $\mathbf{E}_1$. 

\subsection{Staggered $g$-tensor}
\label{gtensordiscussion}
The importance of the staggered $g$-tensor in SrCu$_2$(BO$_3$)$_2$
was pointed out by Miyahara et~al.\cite{Miyahara2004}.
The staggered $g$-tensor exists in SrCu$_2$(BO$_3$)$_2$ because of 
the buckling of Cu-O-B planes below 395\,K.
It  mixes singlet and triplet states similar to the static DM interaction $\mathbf{d_2}$.
The strength of the staggered $g$-tensor interaction can be
estimated and we show that its effect on the energy of the spin levels 
is small compared with the effect of $\mathbf{d_2}$. 
The Zeeman term $H_{Zs}$ couples singlet and triplet states on a single dimer and
is proportional to $g_s\mu_B B_0$,
where $g_s=(g_{\overline{x}}-g_{\overline{z}})\sin\phi\cos\phi$ (Ref.\,\cite{Miyahara2004}).
The angle $\phi\approx 6^\circ$ is the buckling angle of the  Cu-O-B plane\cite{Sparta2001}.
The components $g_{\overline{x}}$ and $g_{\overline{z}}$ of the Cu ion $g$-tensor are not 
known but we take $g_{\overline{x}}\approx g_a=1.998$ 
and $g_{\overline{z}}\approx g_c=2.219$ (Table\,\ref{AllLines})
and get $g_s=0.023$.
The staggered term $H_{Zs}$ increases linearly with magnetic field.
The largest  field where the anti-crossing between $T_0$ and $S_2$
takes place is 5\,T. 
In this field the magnitude of the staggered $g$-tensor term in the hamiltonian
is 0.05\,cm$^{-1}$, which
is much smaller than the static intra-dimer DM term $d_2=1.8$\,cm$^{-1}$.
We conclude that the dominant coupling between the singlet and the triplet 
is due to the static DM interaction $\mathbf{d_2}$.

\subsection{States of bound triplets}

Several states besides the one-triplet excitation 
are infrared-active (Table\,\ref{AllLines}).
We showed that the two dimer model explains
well the energies of the one-triplet states and transitions to them.
In the two dimer model with  $j_1=24$\,cm$^{-1}$ and $2j_2=9.8$\,cm$^{-1}$ 
we get several two triplet states:
a singlet, a triplet, and a quintet of two bound triplets at 28.4, 38.2,
and 57.8\,cm$^{-1}$, respectively. 

SrCu$_2$(BO$_3$)$_2$ has two low energy singlet states 
$S_1$ and $\mathrm {S}_2$
which both anti-cross with triplet state levels
(Fig.\,\ref{BacEcLevelsProbab} and \ref{EparAEigenProb}).
In the two dimer model only one singlet of bound triplets is possible
and the anti-crossing occurs only with $T_{0p}(\pm)$ states.
In the experiment an anti-crossing is observed between
$S_2$ and $T_{0p}(+)$, Fig.\,\ref{EparAEigenProb}(a).
The observed anti-crossing between $S_1$ and $T_{0m}(-)$
cannot be explained by the two dimer model.
In Section\,\ref{DMdiscussion} we show that other
DM interactions besides $\mathbf{d}_2$ are weak
or absent in SrCu$_2$(BO$_3$)$_2$ in the studied $\mathbf{B}_0$ range, 
although they may have proper symmetry
to couple $S_1$ and $T_{0m}(-)$.

The energy of the 38.2\,cm$^{-1}$ triplet in the two dimer model is in
the range where triplets are present in SrCu$_2$(BO$_3$)$_2$.
There is a triplet at 37.5\,cm$^{-1}$ labelled as $T_1$ (Table\,\ref{AllLines}).
FIR transitions to this state are active in $\mathbf{E}_1 \parallel a $ polarization,
Fig.\,\ref{EaEc4K4spec}. 
In $\mathbf{E}_1 \parallel c $ the transitions are weaker (Table\,\ref{AllLines}).
The $T_1(0)$ level is FIR active when  $\mathbf{B}_0 \parallel c $ 
and $T_1(\pm)$ are active when $\mathbf{B}_0 \perp c $.
All this, polarization and magnetic field dependence,
is consistent with the dynamic DM mechanism of the FIR absorption
where the dynamic DM is along the $c$ axis.
The $a$-axis phonon creates a dynamic DM in the direction parallel to the $c$ axis.
The intra-dimer dynamic DM interaction  $d_{3a}$ does not give any transitions
to bound states of triplets.
We considered a possibility that the $a$-axis phonon modulates 
the static inter-dimer $\mathbf{d}_1$.
We found that the pattern of dynamic inter-dimer DM vectors with 
the same symmetry as $\mathbf{d}_1$  [Fig.\,\ref{Model2dimersStaticW}(b)]
gives selection rules that apply to the 37.5\,cm$^{-1}$  $T_1$ triplet.
Transitions to other states are forbidden in the first order of 
this dynamic DM interaction.
The lattice deformation
that creates such a pattern of dynamic DM vectors is of $A_1$ symmetry 
and  is not an optical phonon;
in the $A_1$ symmetry mode Cu atoms on $j_1$ bond move along the bond in antiphase .
We conclude that the two dimer model is not sufficient to account for transitions
to states of bound triplets, except to $S_2	$.

Quintet states were observed by high field ESR.\cite{Nojiri2003}
Their extrapolated zero field energies are in the range 46 - 58\,cm$^{-1}$.
There are two $\mathbf{E}_1 \parallel \mathbf{a}$ 
singlet excitations at 52.3 and 53.5\,cm$^{-1}$ in this range (Table\,\ref{AllLines}).
The quintet ($S=2$) has a $m_S=0$ spin level which has the same 
magnetic field dependence of energy as the $S=0$ state.
However, the observed singlets at 52.3 and 53.5\,cm$^{-1}$  are not the $m_S=0$
components of the quintet.
If in one $\mathbf{B}_0$ field orientation the $m_S=0$
level is infrared-active then in the $90^\circ$ rotated
field orientation other levels, $m_S=\pm 1$ or $m_S=\pm 2$,  
become active.
We  studied all possible $\mathbf{B}_0$, $\mathbf{E}_1$ orientations 
relative to crystal axes 
and did not find the splitting of the 52.3 and 53.5\,cm$^{-1}$    
excitations in the magnetic field although they are one to two 
orders of magnitude more intensive than other magnetic excitations
in FIR spectra.

We assigned the  $a$ axis polarized ($\mathbf{E}_1 \parallel \mathbf{a}$)
43.0, 52.3, and 53.5\,cm$^{-1}$ singlet excitations  to magnetic excitations 
because of the magnetic  field and temperature dependence 
of their energy and intensity.\cite{Room2000}
Whether they could be phonons activated by magnetic interactions  needs a further study.

\section{Conclusions}
 
In SrCu$_2$(BO$_3$)$_2$ the ground state is not  exactly a product of singlets on
dimers as in the Shastry-Sutherland model, because 
the intra-dimer Dzyaloshinskii-Moriya interaction $\mathbf{d}_2$ 
mixes the ground singlet state with the triplet.
From the observed anti-crossing between $T_{0p}(+)$ and $S_2$
we get $d_2=1.8$\,cm$^{-1}$.
This is comparable to the inter-dimer DM, $d_1=0.7$\,cm$^{-1}$, 
which determines the triplet state energy level zero field splitting.
Both  $\mathbf{d}_1$ and $\mathbf{d}_2$ determine 
the magnetic field dependence of the absorption line intensities.

Although  magnetic dipole singlet-triplet transitions are allowed
by $\mathbf{d}_2$, the experimentally observed
polarization and magnetic field dependencies of absorption line intensities
are  not described  by this interaction.   
Instead,  singlet-triplet transitions are allowed by the dynamic 
DM mechanism where  the electric field component of FIR light 
couples to a non-symmetric phonon, which creates the DM interaction.
There are two dynamic DM mechanisms in SrCu$_2$(BO$_3$)$_2$. 
In one case the FIR light couples to an $a$-axis phonon 
and in the other case to a $c$-axis phonon.
This is consistent with the calculations of C{\'e}pas and Ziman\cite{Cepas2004PRB}
who used a two dimer model in the $j_2=0$ limit.

The experiment also yielded information about higher triplet and singlet 
excitations.
Several of these absorption lines are identified as 
electric dipole transitions.
The two dimer cluster is too small 
to describe these transitions.
Also, we had to use renormalized values of $j_1$ and $j_2$
to calculate the energy levels
because the actual spin excitations are delocalized over a larger cluster.
Obviously a bigger cluster is needed for proper
calculation of magnetic excitations in SrCu$_2$(BO$_3$)$_2$.
Nevertheless, the two dimer model gives us a good 
description of the one-triplet excitation.

\section{Acknowledgments}

We thank G.\,Blumberg and O.\,C{\'e}pas for fruitful discussions.
This work was supported by the 
Estonian Science Foundation Grants Nos. 4926, 4927, and 5553.

\bibliographystyle{apsrev}


\begin{thebibliography}{30}
\bibitem[{mis()}]{misalignement}
\bibinfo{note}{Weak 52.3 and 53.5\,cm$^{-1}$ lines are present in $\mathbf{E}_1
  \parallel \mathbf{c}$ spectra because the crystal was misaligned relative to
  the light $\mathbf{k}$ vector; that created a small in-plane $\mathbf{E}_1$
  component. As estimated from the residual intensity of the 52.3 and
  53.5\,cm$^{-1}$ lines in $\mathbf{E}_1 \parallel \mathbf{c}$ polarization the
  $\mathbf{k}$ vector was out of the $(ab)$ plane by $12\pm2^\circ$. The same
  applies to the $\mathbf{B}_0$ field in Faraday configuration,
  $\mathbf{B}_0\parallel \mathbf{k}$. Misalignement of $9\pm1^\circ$ is
  calculated from the splitting of the triplet components,
  Fig.\,\ref{EaBbyEigenProb}.}
\end{thebibliography}

\newpage


\begin{widetext}
\clearpage
\begin{figure}[tb]
    \includegraphics[width=17.8cm]{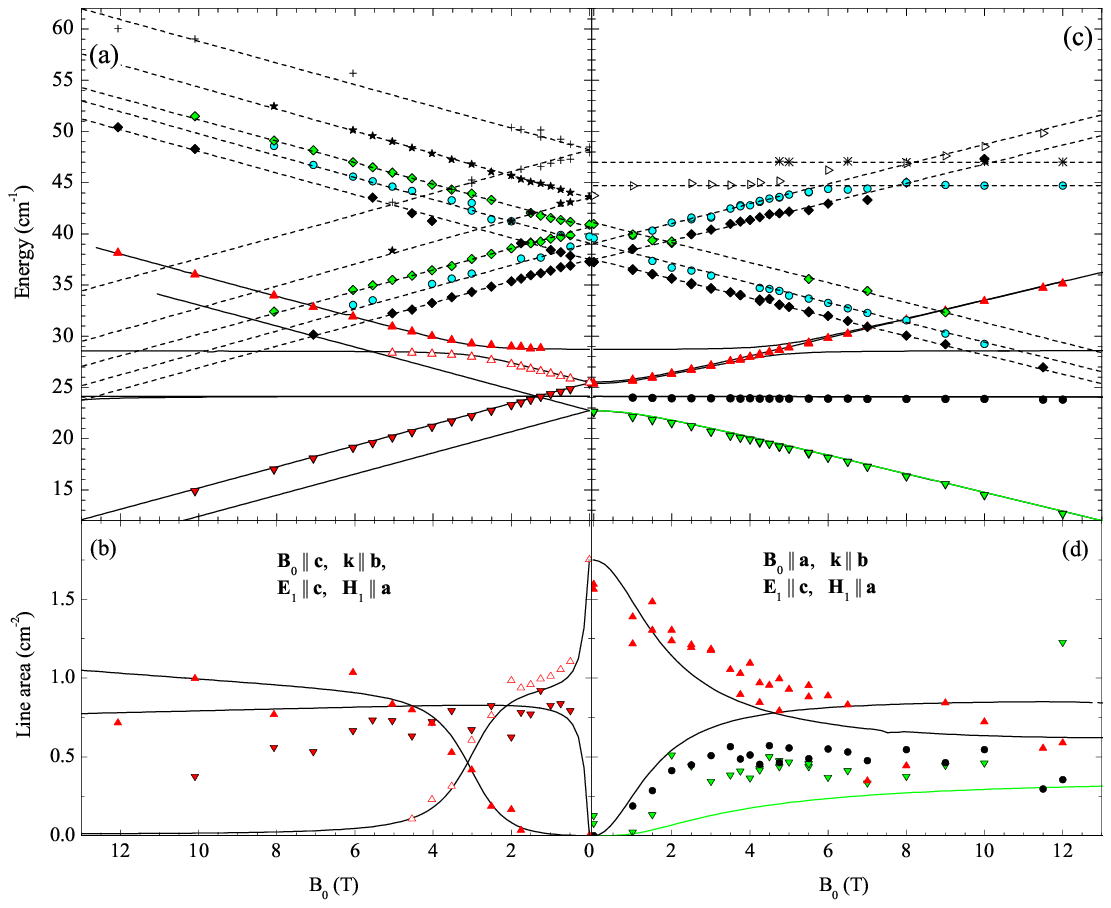}
\caption{(color online).
Magnetic field dependence of line positions and line areas in 
$\mathbf{E}_1 \! \parallel \! \mathbf{c}$ polarization
at 4.4K; (a), (b)  $\mathbf{B}_0 \! \parallel \! \mathbf{c}$;
 (c), (d)  $\mathbf{B}_0 \! \parallel \! \mathbf{a}$. 
Solid lines are the results of the calculation based on the two dimer model: 
$j_1=$24\,cm$^{-1}$, $2j_2=$9.8\,cm$^{-1}$, $2d_1=$1.4\,cm$^{-1}$, and $d_2=$1.8\,cm$^{-1}$.
Dashed lines in panels (a) and (c) are fits with parameters given in Table\,\ref{AllLines}.
}
\label{BacEcLevelsProbab}
\end{figure}

\begin{figure}[tb]
    \includegraphics[width=17.8cm]{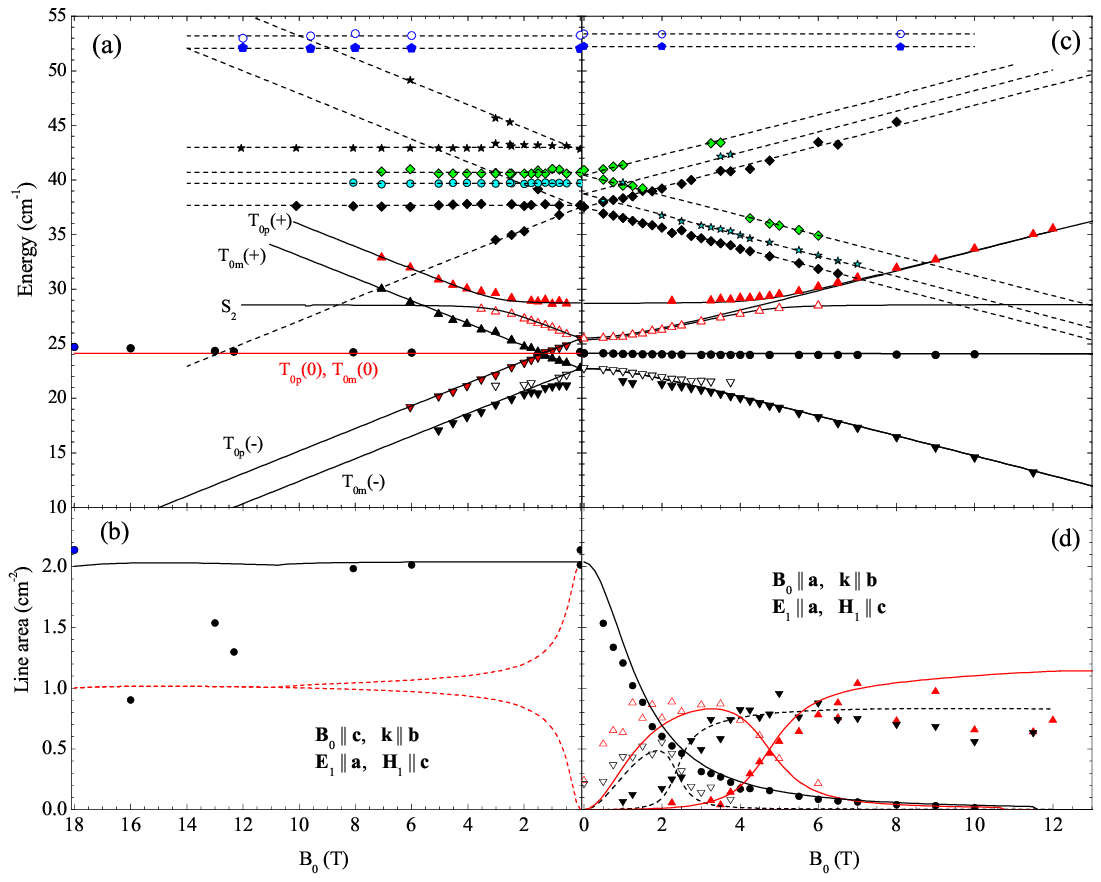}
\caption{(color online).
Magnetic field dependence of line positions and line areas in 
$\mathbf{E}_1 \! \parallel \! \mathbf{a}$ polarization
at 4.4K; (a), (b)  $\mathbf{B}_0 \! \parallel \! \mathbf{c}$;
 (c), (d)  $\mathbf{B}_0 \! \parallel \! \mathbf{a}$. 
Solid lines are the results of the calculation based on the two dimer model:
$j_1=$24\,cm$^{-1}$, $2j_2=$9.8\,cm$^{-1}$, $2d_1=$1.4\,cm$^{-1}$, and $d_2=$1.8\,cm$^{-1}$.
Dashed lines in panels (a) and (c) are fits with parameters given in Table\,\ref{AllLines}.
The solid  line in panel (b) is the sum of two theoretical line areas 
of $S_0$ to $T_{0m}(0)$ and to $T_{0p}(0)$ transitions
shown by dashed lines.
Dashed lines in  (d) are eye guides (see text).
The 18\,T point in panels (a) and (b)  was measured at 1.8\,K.}
\label{EparAEigenProb}
\end{figure}

\clearpage
\end{widetext}

\end{document}